\documentclass[twocolumn,showpacs,aps,prl]{revtex4-1}
\usepackage{amssymb,amsmath}
\usepackage[dvips]{graphicx}
\usepackage[dvipdfm,colorlinks,breaklinks=true,linkcolor=blue,citecolor=blue,urlcolor=blue]{hyperref}

\begin{document}

\title{Phase transitions on random lattices: How random is topological disorder?}

\author{Hatem Barghathi}

\author{Thomas Vojta}
\affiliation{Department of Physics, Missouri University of Science and Technology, Rolla, MO 65409, USA}

\begin{abstract}
We study the effects of topological (connectivity) disorder on phase transitions.
We identify a broad class of random lattices whose disorder fluctuations decay much faster with
increasing length scale than those of generic random systems, yielding a wandering exponent of $\omega=(d-1)/(2d)$
in $d$ dimensions.
The stability of clean critical points is thus
governed by the criterion $(d+1)\nu > 2$ rather than the usual Harris criterion $d\nu>2$, making topological
disorder less relevant
than generic randomness. The Imry-Ma criterion is also modified, allowing first-order transitions
to survive in all dimensions $d>1$. These results explain a host of puzzling violations of the original
criteria for equilibrium and nonequilibrium phase transitions on random lattices. We discuss applications, and we illustrate our theory
by computer simulations of random Voronoi and other lattices.
\end{abstract}

\date{\today}

\maketitle


Two of the central results on phase transitions in disordered systems are
the Harris and Imry-Ma criteria.
The Harris criterion \cite{Harris74} governs the stability of critical points
against disorder. If the correlation length exponent $\nu$ of
a $d$-dimensional clean system fulfills the inequality $d\nu>2$,
weak disorder is irrelevant and does not change the critical behavior.
If $d\nu<2$, disorder is relevant, and the character of the transition must change
\cite{Vojta06,*Vojta10}.
The Imry-Ma criterion \cite{ImryMa75,*ImryWortis79,*HuiBerker89,*AizenmanWehr89} governs the stability
of macroscopic phase coexistence: Disorder destroys phase coexistence
by domain formation in dimensions $d\le 2$
\footnote{If the randomness breaks a continuous symmetry, the marginal
dimension is $d = 4$.}.
As a consequence, disorder rounds first-order phase transitions in
$d\le 2$.
The predictions of these criteria and their generalizations
to long-range correlated disorder \cite{WeinribHalperin83,Nattermann83} agree
with the vast majority of explicit results on classical, quantum, and
nonequilibrium  systems in which the disorder stems
from random coupling strengths or spatial dilution.

Puzzling results have been reported, however, on phase
transitions in \emph{topologically disordered} systems, i.e., systems on lattices
with random connectivity.
For example, the Ising magnet on a three-dimensional (3D) random Voronoi lattice displays the same
critical behavior as the Ising model on a cubic lattice \cite{JankeVillanova02,LimaCostaCosta08}
even though Harris' inequality is violated. An
analogous violation was found for the 3-state Potts model on a 2D random
Voronoi lattice \cite{LCAA00}. The regular 2D 8-state Potts model
features a first-order phase transition. In contrast to the prediction of the
Imry-Ma criterion, the transition remains of first order on a random Voronoi
lattice \cite{JankeVillanova95}.

The nonequilibrium transition of the contact process features an even more striking discrepancy.
This system violates Harris' inequality \cite{Hinrichsen00,*Odor04}.
Disorder introduced via dilution or random transition rates
results in an infinite-randomness critical point and
strong Griffiths singularities \cite{HooyberghsIgloiVanderzande03,*HooyberghsIgloiVanderzande04,VojtaDickison05,*VojtaFarquharMast09,*Vojta12}.
In contrast, the contact process on a 2D random Voronoi lattice
shows clean critical behavior and no trace of the exotic strong-randomness
physics \cite{OAFD08}.

To explain the unexpected failures of the Harris and Imry-Ma criteria, several authors
suggested that, perhaps, the existing results are not in the asymptotic regime.
Thus, much larger systems would be necessary to observe the true asymptotic behavior which, presumably,
agrees with the Harris and Imry-Ma criteria. However, given the large systems employed in
some of the cited work, this would imply enormous crossover lengths
which do not appear likely because the coordination number fluctuations
of the Voronoi lattice are not particularly small
\footnote{The standard deviation of the coordination number is about 20\% of its average in both
          2D and 3D.}.
What, then, causes the failure of the Harris and Imry-Ma criteria
on random Voronoi lattices?

In this Letter, we show that 2D random Voronoi lattices belong to
a broad class of random lattices whose disorder fluctuations
feature strong anticorrelations and thus decay qualitatively faster with
increasing length scale than those of generic random systems. This class
comprises lattices whose total coordination (total number of bonds) does
not fluctuate. Such lattices are particularly prevalent in 2D because the
Euler equation of a 2D graph imposes a topological constraint on the coordination
numbers. However, higher-dimensional realizations exist as well.
The suppressed disorder fluctuations lead to an important modification of the
Harris criterion: The random connectivity is irrelevant at
clean critical points if $(d+1)\nu > 2$.
Topological disorder is thus less relevant than generic randomness. The
Imry-Ma criterion is modified as well, allowing first-order transitions to survive
in all dimensions $d>1$.  This explains the puzzling literature results on 2D random Voronoi lattices
mentioned above.
In the rest of this Letter, we sketch the derivation of these results and illustrate them
by simulations.

\begin{figure}
\includegraphics[width=8.2cm]{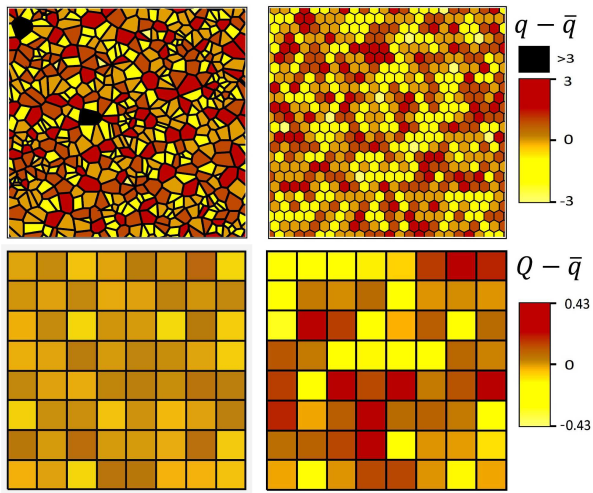}
\caption{(color online). Top row: coordination numbers $q_i$ of individual sites in
      a random Voronoi lattice (left) and a diluted hexagonal lattice (right). Bottom row: average coordination
      number $Q_\mu$ of blocks with $L_b=8$. The strong suppression of the fluctuations in the Voronoi
      lattice is clearly visible. (The same color (gray) scale is used left and right).}
\label{fig:lattices}
\end{figure}

Random lattice or cell structures occur in many areas of physics, chemistry, and biology
such as amorphous solids, foams, and biological tissue. Consider a many-particle system on
such a random  lattice, e.g., a classical or quantum spin system, lattice bosons,
or a nonequilibrium problem such as the contact process.
In all these examples, the disorder of the many-particle system stems from the
random connectivity of the underlying lattice. In the following, we therefore
analyze the fluctuations of the coordination number $q_i$ (the number of nearest
neighbors of site $i$) for different random lattices, starting
with the 2D random Voronoi lattice (Fig.\ \ref{fig:lattices}).

The Voronoi-Delaunay construction is an algorithm for building a cell network from a set of
lattice sites \cite{OBSC_book00}.
The Voronoi cell of a site consists of all points in the plane
that are closer to this site than to any other. Sites whose Voronoi cells
share an edge are considered neighbors. The graph of all bonds connecting
pairs of neighbors defines a triangulation of the plane called the
Delaunay triangulation.
Our simulations start by performing the Voronoi-Delaunay construction
\cite{[{We use an efficient algorithm inspired by }] TanemuraOgawaOgita83}
for $N$ points placed at independent random positions within a square
of side $L=N^{1/2}$ (density fixed at unity). To study the coordination number fluctuations,
we divide the system into square blocks of side $L_b$ and calculate the block-averaged
coordination number
\begin{equation}
Q_\mu = N_{b,\mu}^{-1} \sum_{i\in \mu} q_i
\label{eq:block_Q}
\end{equation}
for each block. $N_{b,\mu}$ is the number of sites in block $\mu$, and the sum runs over all these sites.
The relevant quantity is the standard deviation $\sigma_Q$ of the block-averaged coordination numbers
defined by
\begin{equation}
\sigma^2_Q (L_b) = \left [ (Q_\mu - \bar q)^2 \right ]_{\mu}
\label{eq:sigma_Q}
\end{equation}
where $[\ldots]_\mu$ denotes the average over all blocks $\mu$,
and $\bar q$ is the global average coordination number of the lattice.

\begin{figure}
  \includegraphics[width=8.2cm]{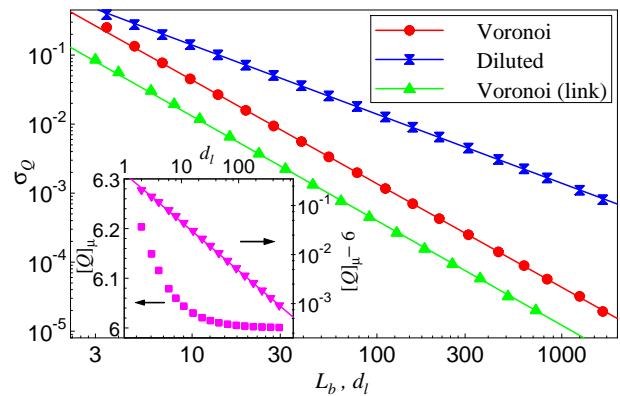}
  \caption{(color online). Standard deviation $\sigma_Q$ of the average
  coordination number $[Q]_\mu$ of blocks of size $L_b$ for a random Voronoi lattice and
  a square lattice with 50\% bond dilution (100 lattices with $5000^2$ sites each).
  The  lines are fits to $\sigma_Q \sim L_b^{-a}$ giving exponents $a=1.001(2)$ (diluted) and $1.501(3)$
  (Voronoi). Also shown is $\sigma_Q$ for
  clusters defined via the link distance $d_l$ (100 lattices with 2000$^2$ sites)
  giving $a=1.52(2)$. Inset: $[Q]_\mu$ and $[Q]_\mu-\bar q$ of the link-distance
  clusters vs.\ $d_l$.  The  line is a fit
  to $([Q]_\mu-\bar q) \sim d_l^{-b}$ yielding $b=0.99(1)$.}
  \label{fig:qfluc2d}
\end{figure}

Figure \ref{fig:qfluc2d} compares the fluctuations in a random Voronoi lattice
and a bond-diluted square lattice (both with periodic boundary conditions).
In the diluted lattice, the fluctuations accurately follow $\sigma_Q \sim L_b^{-d/2}=L_b^{-1}$,
as expected for uncorrelated disorder. In contrast, the fluctuations in
the Voronoi lattice decay faster and follow  $\sigma_Q \sim L_b^{-3/2}$.
 An illustration
of the suppressed fluctuations in the Voronoi lattice is shown in Fig.\ \ref{fig:lattices}.

In addition to real-space blocks, we also study
clusters based on the link distance, the smallest number of
bonds (links) that separate two sites. To construct such clusters, we start from
a random seed site and add its neighbors, neighbors of neighbors and so on until we reach
a maximum link distance $d_l$. This construction introduces a bias towards large $q_i$
(as sites with more neighbors are more likely to be added to the cluster).
Thus, the cluster average $[Q]_\mu$ is larger than the
global average $\bar q=6$, see inset of Fig.\ \ref{fig:qfluc2d}.
The excess decays only slowly with cluster size,
$([Q]_\mu-6) \sim d_l^{-1}$. For the link-distance clusters we therefore use
$\sigma^2_Q (d_l) = \left [ (Q_\mu - [Q]_\mu )^2 \right ]_{\mu}$
rather than eq.\ (\ref{eq:sigma_Q}). The resulting data, also shown in Fig.\ \ref{fig:qfluc2d},
demonstrate that the fluctuations of the link-distance clusters decay
with the same power, $\sigma_Q \sim d_l^{-3/2}$, as those of the real-space blocks. Had we not corrected for the
size-dependence of $[Q]_\mu$, we would have obtained a spurious
decay exponent of $(-1)$
\cite{[{This may explain why an earlier study of the random Voronoi lattice, }] [{, reported
the uncorrelated randomness result $\sigma_Q \sim d_l^{-1}$ }]
JankeWeigel04}.


How can we understand the rapidly decaying disorder fluctuations?
The Euler equation of a 2D graph consisting of $N$ sites, $E$ edges (nearest-neighbor bonds), and
$F$ facets reads
$N-E+F=\chi$. Here, $\chi$ is the Euler characteristic,
a topological invariant of the underlying surface. Periodic boundary conditions
are equivalent to a torus topology, yielding $\chi=0$
\footnote{Other boundary conditions with different values of $\chi$ will just produce
subleading corrections to our results.}.
Every facet of a Delaunay triangulation is a triangle.
As each triangle has three edges, and each edge is shared
by two triangles, $3F=2E$. This implies $E=3N$, i.e., the total coordination does not fluctuate, and the
average coordination number is $\bar q = 2E/N=6$ for any disorder realization.
(This also follows from the angle sum in any triangle being $\pi$: As each site has a total angle of $2\pi$,
6 triangles meet at a site on average.)
Now consider a block of size $L_b$ as introduced above. The relation
$3F=2E$ holds for all triangles and edges completely inside the block.
Any deviation of the block-averaged coordination number $Q_\mu$ from $\bar q = 6$ must thus stem from the block
surface. The number of facets crossing the surface scales linearly with $L_b$. Assuming that each of
these facets makes an independent random contribution to  $Q_\mu$ leads to the estimate
$\sigma_Q (L_b) \sim L_b^{1/2} / L_b^2 = L_b^{-3/2}$ in perfect agreement with the numerical data.

\begin{figure}
\includegraphics[width=8.2cm]{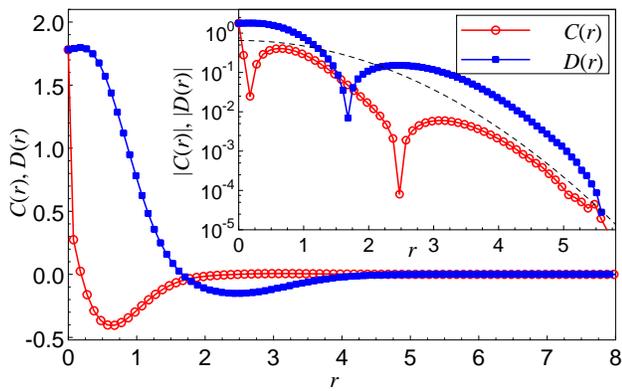}
\caption{(color online). Coordination number correlation function $C(r)$ and its integral $D(r)$
vs.\ distance $r$ averaged over $10^7$ lattices of $24^2$ sites. Inset:
Semi-log plot of $|C(r)|$ and $|D(r)|$. The envelope of $C(r)$ follows
a Gaussian with a characteristic length $x_0 \approx 1.25$ (dashed line). }
\label{fig:corrfunc}
\end{figure}

To substantiate these arguments, we study the coordination number correlation function
\begin{equation}
C(\mathbf{r}) = \frac 1 N \sum_{ij} (q_i - \bar q)(q_j - \bar q) \delta(\mathbf{r}-\mathbf{r}_{ij})
\label{eq:corrfunc}
\end{equation}
where $\mathbf{r}_{ij}$ is the vector from site $i$ to $j$. Its integral over a block
of radius $r$ yields the bulk contribution to the fluctuations of the average coordination number
\begin{equation}
\sigma^2_{Q,\rm bulk}(r) =  D(r)= \frac {2\pi} {N_r} \int_0^{r}  dr' \,r'\, C(r')
\label{eq:C-sigma}
\end{equation}
where $N_r$ is the number of sites in the block.
The data presented in Fig.\ \ref{fig:corrfunc}
show that $|C(r)|$ decays faster than exponential
with distance $r$. Its integral $D(r)$ also decays rapidly to zero, confirming that the total
coordination is not fluctuating. The topological constraint
imposed by the Euler equation thus leads to strong coordination number anticorrelations
that are fully established within 5 or 6 typical nearest-neighbor distances.


How general are these results? Are they restricted to 2D random Voronoi lattices or do they apply
to other lattices as well? The fixed total coordination is a direct consequence of the
Euler equation $N-E+F=\chi$ and the triangle
condition $3F=2E$. It thus applies to any tiling of the plane with triangles.
Analogously, if we tile the plane with arbitrary quadrilaterals, $4F=2E$. This yields a fixed
average coordination number of precisely $\bar q = 2E/N=4$.
We have thus identified a broad class of 2D lattices in which the coordination
fluctuations are suppressed because the total coordination is constrained. In
addition to random Voronoi lattices it includes, e.g., regular lattices with bond-exchange defects
which are related to the topological models of Le Ca\"{e}r \cite{LeCaer91a,*LeCaer91b}.
It also includes deterministic quasiperiodic lattices such as the Penrose and Ammann-Beenker tilings
\cite{GruenbaumShephard86} (using rhombic tiles) as well as random tilings \cite{Henley99}
whose tiles are either all triangles or all quadrilaterals.

What about higher dimensions? The Euler equation for a 3D tessellation, $N-E+F-C=\chi$, contains one extra
degree of freedom, viz., the number $C$ of 3D cells. The total coordination of a random tetrahedralization
is therefore \emph{not} fixed by a topological constraint, in agreement with the fact that the solid-angle
sum in a tetrahedron is \emph{not} a constant. Consequently, 3D random Voronoi lattices do \emph{not}
belong to our class of lattices with a constrained total coordination. However, 3D members of our class
do exist. They include, e.g., lattices built exclusively from rhombohedra such as the icosahedral tiling and
its random variants \cite{ShawElserHenley91} (the solid angle sum of a rhombohedron is fixed at 4$\pi$)
as well as generalizations of the bond-exchange lattices to 3D.


We now generalize to arbitrary dimension our estimate of the fluctuations of the block-averaged
coordination number. As the bulk contribution is suppressed by the anticorrelations,
the main contribution stems from the surface. The number of cells or facets close
to the surface scales as $L_b^{d-1}$ with block-size $L_b$. In the generic case, i.e., in the
absence of further constraints or long-range correlations, these surface cells make independent
random contributions to $Q_\mu$. This leads to
\begin{equation}
\sigma_Q (L_b) \sim L_b^{(d-1)/2} / L_b^d = L_b^{-(d+1)/2}.
\label{eq:sigma_estimate}
\end{equation}
Casting this result in terms of the wandering exponent $\omega$ defined via $\sigma_Q \sim L_b^{-d(1-\omega)}$
\cite{Luck93a}, we obtain $\omega=(d-1)/(2d)$. This needs to be compared to
uncorrelated randomness for which $\sigma_Q \sim L_b^{-d/2}$ and $\omega=1/2$.

\begin{figure}
\includegraphics[width=8.5cm]{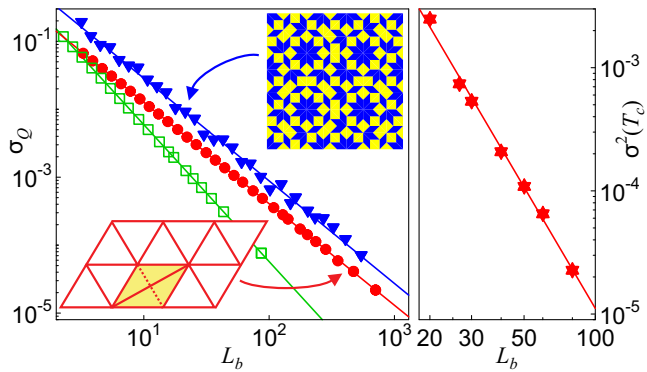}
\caption{(color online). Left: $\sigma_Q$ vs.\ $L_b$ for the Ammann-Beenker
     tiling (8th generation, 6430084 sites, triangles), a triangular lattice with 50\%
     bond-exchange defects (100 lattices with $2000^2$ sites, circles), and
     a rhombohedral lattice with 50\%
     bond-exchange defects (100 lattices with $300^3$ sites, open squares).
     The lines are power-law fits giving exponents of 1.51(3), 1.498(2), and
     2.01(1), respectively.
     Right: $\sigma^2(T_c)$ vs.\ $L_b$ for
     an Ising model on a random Voronoi lattice
     (100 lattices of $100^2$ sites, $10^5$ Monte Carlo sweeps each). The  line is
     a fit to  $\sigma(T_c) \sim L_b^{-c}$ giving $c = 1.56(7)$.}
\label{fig:q_other}
\end{figure}

We have verified the prediction (\ref{eq:sigma_estimate}) for several lattices in addition to the 2D Voronoi lattice.
The first is a random lattice produced from a triangular lattice by performing
random bond exchanges. A bond exchange (left inset of Fig.\ \ref{fig:q_other}) consists in randomly choosing a rhombus
made up of two adjacent triangles and replacing the short diagonal (dotted)  with the long one (solid).
The second example is the deterministic quasiperiodic Ammann-Beenker tiling. For both lattices,
the numerical data (Fig.\ \ref{fig:q_other}) follow $\sigma_Q(L_b) \sim L_b^{-3/2}$ in agreement with (\ref{eq:sigma_estimate})
\footnote{The results for the Ammann-Beenker tiling were obtained using generic random orientations
of the blocks. If square blocks are oriented parallel to the symmetry axes of the tiling, the surface
terms in this \emph{nonrandom} lattice add constructively leading to $\sigma_Q (L_b) \sim L_b^{-1}$.}.
Finally, we have studied a 3D rhombohedral lattice with bond-exchange defects. The numerical data
are in excellent agreement with the prediction $\sigma_Q (L_b) \sim L_b^{-2}$.


We now use these results to derive the analog of the Harris criterion for many-particle
systems on random lattices in our class. Following Harris and Luck \cite{Harris74,Luck93a},
we compare the fluctuations of the local distance from criticality between correlation volumes
with the global distance from criticality. If the interactions between the sites
are restricted to nearest neighbors and of equal strength, the disorder fluctuations
are governed by (\ref{eq:sigma_estimate}) and decay as
$\xi^{-(d+1)/2}$ with correlation length $\xi$. The global distance from criticality scales as
$\xi^{-1/\nu}$. A clean critical point is thus stable if $\xi^{-(d+1)/2} < \xi^{-1/\nu}$
for $\xi \to \infty$. This yields the stability (Harris-Luck) criterion $(d+1)\nu > 2$.
The topological disorder is thus less relevant than generic uncorrelated randomness
for which the Harris criterion reads $d\nu >2$.

The Imry-Ma criterion compares the free energy gain due to forming a domain that takes
advantage of a disorder fluctuation with the energy cost of the domain wall.
In our class of lattices, the gain scales as $L_b^{d\omega}=L_b^{(d-1)/2}$ while the cost of a
domain wall scales as $L_b^{d-1}$. Forming large domains is thus unfavorable in all
dimensions $d>1$ implying that first-order transitions can survive.

The coordination number fluctuations determine the bare (in the renormalization group sense)
disorder of the many-particle system. To study an example of disorder renormalizations, we calculate the local
critical temperatures $T_c$ of the Ising model, $H=-J \sum_{\langle ij\rangle} S_i S_j$, on a random Voronoi lattice by  Monte-Carlo
simulations.
The right panel of Fig.\ \ref{fig:q_other} shows the variance of the block $T_c$ (defined as the maximum of the susceptibility)
as a function of block size.
The data follow $\sigma(T_c) \sim L_b^{-3/2}$ in agreement with the coordination number. In general, disorder
renormalizations can be expected to generate weak uncorrelated disorder even if the bare disorder is anticorrelated \cite{SVSN93}.
Our results suggest that this uncorrelated disorder, if any, is very weak (as it is invisible on length scales below $L_b \approx 100$)
and thus unobservable in most experiments and simulations.


In summary, we have studied the effects of topological disorder on phase transitions. We have identified a broad class of random lattices
characterized by strong disorder anticorrelations. Such lattices are ubiquitous in 2D because the Euler equation
imposes a topological constraint on the coordination numbers. However, we have also found higher-dimensional realizations.
The anticorrelations lead to modifications of the Harris and Imry-Ma criteria.
This explains most of the puzzling apparent failures of the usual criteria discussed in the introduction.
Note that another type of anticorrelations was recently found to protect a clean critical point in a quantum spin chain \cite{HLVV11}.
Moreover, local disorder correlations that change the degree of frustration in a spin glass can qualitatively change
its phase diagram \cite{IlkerBerker14}.

Interestingly, the 3D random Voronoi lattice does not belong to our class of lattices with constraint total
coordination. Preliminary numerical results suggest that its coordination number fluctuations decay
more slowly than (\ref{eq:sigma_estimate}) but still faster than the uncorrelated randomness result $L_b^{-d/2}$,
at least for blocks with $L_b < 400$. Further work will be necessary to understand the fate of phase transitions
on 3D Voronoi lattices.

So far, we have considered systems in which all pairs of neighbors interact equally strongly. If this is not so,
e.g., because the interactions depend on the distance between neighboring sites, the disorder anticorrelations are
destroyed. The critical behavior is thus expected to cross over to that of uncorrelated disorder. We have explicitly observed this crossover
in the contact process \cite{BarghathiVojta_unpublished}.

It will be interesting to study transitions that violate even the modified stability criterion $(d+1)\nu > 2$.  A prime example is the
quantum phase transition of the transverse-field Ising magnet on a 2D random Voronoi lattice. Its clean critical behavior is in the (2+1)D
Ising universality class with $\nu\approx 0.630$ and thus violates $(d+1)\nu > 2$. As the anticorrelations strongly suppress
the rare region probability \cite{BarghathiVojta_unpublished}, we also expect significant modifications of the quantum Griffiths singularities.


This work was supported by the NSF under Grant Nos.\ DMR-1205803 and PHYS-1066293.
We acknowledge the hospitality of the Aspen Center for Physics.

\bibliographystyle{apsrev4-1}
\bibliography{../00Bibtex/rareregions}

\end{document}